\documentclass[aps,prb,floatfix,twocolumn,showpacs]{revtex4}
\usepackage{amsmath}
\usepackage{graphicx}
\usepackage{amsfonts}
\usepackage{amssymb}
\begin{document}

\title{Spectroscopy and critical temperature of diffusive
superconducting/ferromagnetic hybrid structures with spin-active interfaces}
\author{Audrey Cottet}
\affiliation{Laboratoire de Physique Th\'{e}orique et Hautes \'{E}nergies, Universit\'{e}s Paris 6 et 7, CNRS, UMR 7589, 4 place Jussieu, F-75252 Paris Cedex 05, France}
\affiliation{Laboratoire de Physique des Solides, Universit\'{e} Paris-Sud, CNRS, UMR 8502,  
F-91405 Orsay Cedex, France}
\begin{abstract}
The description of the proximity effect in
superconducting/ferromagnetic heterostructures requires to use spin-dependent
boundary conditions. Such boundary conditions must take into account the spin
dependence of the phase shifts acquired by electrons upon scattering on the
boundaries of ferromagnets. The present article shows that this property can
strongly affect the critical temperature and the energy dependence of the
density of states of diffusive heterostructures. These effects should allow a
better caracterisation of diffusive superconductor/ferromagnet interfaces.
\end{abstract}
\date{\today}
\maketitle

\section{Introduction}

When a ferromagnetic metal ($F$) with uniform magnetization is connected to a
BCS superconductor ($S$), the singlet electronic correlations characteristic
of the $S$ phase can propagate into $F$ because electrons and holes with
opposite spins and excitation energies are coupled coherently by Andreev
reflections occurring at the $S/F$ interface. Remarkably, the ferromagnetic
exchange field induces an energy shift between the coupled electrons and
holes, which leads to spatial oscillations of the superconducting order
parameter in $F$ \cite{Buzdin1982,Golubov}. This effect has been observed
experimentally through oscillations of the density of states (DOS) in $F$ with
the thickness of $F$ \cite{TakisN}, or oscillations of the critical current
$I_{0}$ through $S/F/S$ structures \cite{Ryazanov,TakisI,SellierPRB,Blum},
with the thickness of $F$ or the temperature. The oscillations of $I_{0}$ have
allowed to obtain $\pi$-junctions\cite{Guichard}, i.e. Josephson junctions
with $I_{0}<0$, which could be useful in the field of superconducting circuits
\cite{Ioffe,Taro}. A reentrant behavior of the superconducting critical
temperature of $S/F$ bilayers with the thickness of $F$ has also been observed
\cite{TcSF}. At last, some $F/S/F$ trilayers have shown a lower critical
temperature for an antiparallel alignment of the magnetizations in the two $F$
layers as compared with the parallel alignment\cite{SSspinswitch}, which
should offer the possibility of realizing a superconducting
spin-switch\cite{DeGennes,TcFSFth}. \begin{figure}[ptb]
\includegraphics[width=0.7\linewidth]{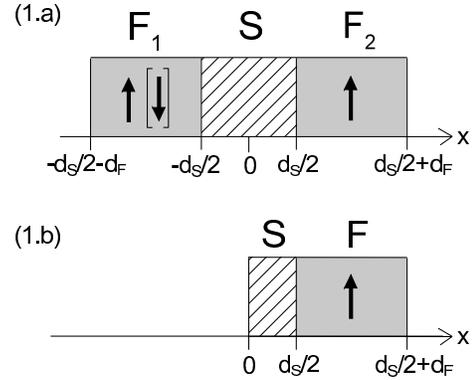}\caption{a. Diffusive
$F/S/F${\LARGE \ }trilayer consisting of a BCS\ superconductor $S$ with
thickness $d_{S}$ placed between two ferromagnetic electrodes $F_{1}$ and
$F_{2}$ with thickness $d_{F}$. In this picture, the directions of the
magnetic polarizations in $F_{1}$ and $F_{2}$ \ are parallel [antiparallel],
which corresponds to the configuration $\mathcal{C}=P$ $[AP]$. b.
$S/F${\LARGE \ }bilayer consisting of a BCS\ superconductor $S$ with thickness
$d_{S}/2$ contacted to a ferromagnetic electrode $F$ with thickness $d_{F}$.}%
\end{figure}

For a theoretical understanding of the behavior of $S/F$ hybrid circuits, a
proper description of the interfaces between the different materials is
crucial. For a long time, the only boundary conditions available in the
diffusive case were spin-independent boundary conditions derived for
$S/$normal metal interfaces\cite{Kuprianov}. Recently, spin-dependent boundary
conditions have been introduced for describing hybrid diffusive circuits
combining BCS superconductors, normal metals and ferromagnetic insulators
\cite{condmatHuertas}. These boundary conditions take into account the
spin-polarization of the electronic transmission probabilities through the
interface considered, but also the spin-dependence of the phase shifts
acquired by electrons upon transmission or reflection by the interface. The
first property generates widely known magnetoresistance effects\cite{magn}.
The second property is less commonly taken into account. However, the
Spin-Dependence of Interfacial Phase Shifts (SDIPS) can modify the behavior of
many different types of mesoscopic circuits with ferromagnetic elements, like
those including a diffusive normal metal island \cite{FNF}, a resonant
system\cite{CottetEurophys,SST}, a Coulomb blockade system\cite{CB,
Cottet06,SST}, or a Luttinger liquid\cite{Luttinger}. It has also been shown
that the SDIPS has physical consequences in $S/F$ hybrid
systems\cite{Tokuyasu,otherBC,mixing,condmatHuertas}. One can note that, in
some references, the SDIPS is called ''spin-mixing angle'' or ''spin-rotation
angle'' (see e.g. Refs. \onlinecite{Tokuyasu, mixing}).\textbf{\ }In the
diffusive $S/F$ case, the spin-dependent boundary conditions of
Ref.~\onlinecite
{condmatHuertas} have been applied to different circuit
geometries\cite{demoBC,applications,Morten,cottet05,Braude} but the only
comparison to experimental data has been performed in Ref.~\onlinecite
{cottet05}. The authors of this Ref. have generalized the boundary conditions
of Ref.~\onlinecite{condmatHuertas} to the case of metallic $S/F$ interfaces
with a superconducting proximity effect in $F$. They have showed that the
SDIPS can induce a shift in the oscillations of the critical current of a
$S/F/S$ Josephson junction or of the DOS of a $S/F$ bilayer with the thickness
of $F$. Signatures of this effect have been identified in the \textrm{Nb/PdNi}
hybrid structures of Refs.~\onlinecite{TakisN,TakisI}. Nevertheless, the
problem of characterizing the SDIPS of diffusive $S/F$ interfaces has raised
little attention so far, in spite of the numerous experiments performed.

A good characterization of the properties of diffusive $S/F$ interfaces would
be necessary for a better control of the superconducting proximity effect in
diffusive heterostructures. The present article presents other consequences of
the SDIPS than that studied in Ref.~\onlinecite{cottet05}, which could be
useful in this context. In particular, the SDIPS can generate an effective
magnetic field in a diffusive $S$ in contact with a diffusive $F $, like found
for a ballistic $S$ in contact with a ferromagnetic insulator \cite{Tokuyasu}.
This effective field can be detected, in particular, through the DOS of the
diffusive $F$ layer, with a visibility which depends on the thickness of $F$.
A strong modification of the variations of the critical temperature of
diffusive $S/F$ structures with the thickness of $F$ is also found. These
effects should allow to characterize the SDIPS of diffusive $S/F$ interfaces
through DOS and critical temperature measurements, by using the
heterostructures currently fabricated. The calculations reported in this paper
are also appropriate to the case of a diffusive $S$ layer contacted to a
ferromagnetic insulator ($FI $).

This paper is organized as follows: Section II presents the initial set of
equations used to describe the heterostructures considered. The case of
$F/S/F$ trilayers is mainly addressed, but the case of $S/F$ (or $S/FI$)
bilayers follows straightforwardly. Section III specializes to the case of a
weak proximity effect in $F$ and a superconducting layer with a relatively low
thickness $d_{S}\leq\xi_{S}$, with $\xi_{S}$ the superconducting coherence
length in $S$. The spatial evolution of the electronic correlations in the $S$
and $F$ layers is studied in Section III.A. The energy-dependent DOS of $S/F$
heterostructures is calculated in Section III.B. Section III.C considers
briefly the limit of $S/FI$ bilayers. Section III.D discusses SDIPS-induced
effective field effects in other types of systems. Section III.E compares the
present work to other DOS calculations for data interpretation in $S/F$
heterostructures. Critical temperatures of $S/F$ circuits are calculated and
discussed in Section III.F. Conclusions are presented in Section IV.
Throughout the paper, I consider conventional BCS superconductors with a
s-wave symmetry.

\section{Initial description of the problem}

This article mainly considers a diffusive $F/S/F${\LARGE \ }trilayer
consisting of a BCS\ superconductor $S$ for $-d_{S}/2<x<d_{S}/2$, and
ferromagnetic electrodes $F_{1}$ for $x\in\{-d_{S}/2-d_{F},-d_{S}/2\}$ and
$F_{2}$ for $x\in\{d_{S}/2,d_{S}/2+d_{F}\}$ (see Figure 1.a). The magnetic
polarization of the two ferromagnets can be parallel (configuration
$\mathcal{C}=P$) or anti-parallel (configuration $\mathcal{C}=AP$), but the
modulus $\left|  E_{ex}\right|  $ of the ferromagnetic exchange field is
assumed to be the same in $F_{1}$ and $F_{2}$. Throughout the structure, the
normal quasiparticle excitations and the superconducting condensate of pairs
can be characterized with Usadel normal and anomalous Green's functions
$G_{n,\sigma}=\mathrm{sgn}(\omega_{n})\cos(\theta_{n,\sigma})$ and
$F_{n,\sigma}=\sin(\theta_{n,\sigma})$, with $\theta_{n,\sigma}(x)$ the
superconducting pairing angle, which depends on the spin direction $\sigma
\in\{\uparrow,\downarrow\}$, the Matsubara frequency $\omega_{n}(T)=(2n+1)\pi
k_{B}T$, and the coordinate $x$ (see e.g. Ref.~\onlinecite{RevueW}). The
Usadel equation describing the spatial evolution of $\theta_{n,\sigma}$ writes%

\begin{equation}
\frac{\hbar D_{S}}{2}\frac{\partial^{2}\theta_{n,\sigma}}{\partial x^{2}%
}=\left|  \omega_{n}\right|  \sin(\theta_{n,\sigma})-\Delta(x)\cos
(\theta_{n,\sigma}) \label{UsadelS}%
\end{equation}
in $S$ and
\begin{equation}
\frac{\hbar D_{F}}{2}\frac{\partial^{2}\theta_{n,\sigma}}{\partial x^{2}%
}=\left(  \left|  \omega_{n}\right|  +iE_{ex}\sigma\mathrm{sgn}(\omega
_{n})\right)  \sin(\theta_{n,\sigma}) \label{UsadelF2}%
\end{equation}
in $F_{1}$ and $F_{2}$, with $D_{F}$ the diffusion constant of the
ferromagnets and $D_{S}$ the diffusion constant of $S$. The self-consistent
superconducting gap $\Delta(x)$ occurring in (\ref{UsadelS}) can be expressed
as
\begin{equation}
\Delta(x)=\frac{\pi k_{B}T\lambda}{2}\sum\limits_{\substack{\sigma
\in\{\uparrow,\downarrow\}\\\omega_{n}(T)\in\{-\Omega_{D},\Omega_{D}\}}%
}\sin(\theta_{n,\sigma}) \label{Delta}%
\end{equation}
with $\Omega_{D}$ the Debye frequency of $S$, $\lambda^{-1}=2\pi k_{B}T_{_{c}%
}^{BCS}\sum\nolimits_{\omega_{n}(T_{_{c}}^{BCS})\in\{0,\Omega_{D}\}}\omega
_{n}^{-1}$ the BCS coupling constant and $T_{_{c}}^{BCS}$ the bulk transition
temperature of $S$. I assume $\Delta=0$ in $F_{1}$ and $F_{2}$. The above
equations must be supplemented with boundary conditions describing the
interfaces between the different materials. First, one can use
\begin{equation}
\left.  \frac{\partial\theta_{n,\sigma}}{\partial x}\right|  _{x=\pm
(d_{S}/2+d_{F})}=0 \label{derZero}%
\end{equation}
for the external sides of the structure. Secondly, the boundary conditions at
the $S/F$ interfaces can be calculated by assuming that the interface
potential locally dominates the Hamiltonian, i.e. at a short distance it
causes only ordinary scattering (with no particle-hole mixing) (see e.g.
Ref.~\onlinecite{Nazarov}). This ordinary scattering can be described with
transmission and reflection amplitudes $t_{n,\sigma}^{S(F)}$ and $r_{n,\sigma
}^{S(F)}$ for electrons coming from the $S$($F$) side of the interface in
channel $n$ with a spin direction $\sigma$. The phases of $t_{n,\sigma}%
^{S(F)}$ and $r_{n,\sigma}^{S(F)}$ can be spin-dependent due to the exchange
field $E_{ex}$ in $F_{1(2)}$ and a possible spin-dependence of the barrier
potential between $S$ and $F_{1(2)}$. Boundary conditions taking into account
this so-called Spin-Dependence of Interfacial Phase Shifts (SDIPS) have been
derived for $|t_{n,\uparrow}^{S}|^{2},|t_{n,\downarrow}^{S}|^{2}\ll1$ and a
weakly polarized $F$\cite{condmatHuertas,cottet05}. When there is no SDIPS,
the boundary conditions involve the tunnel conductance $G_{T}=G_{Q}%
\sum\nolimits_{n}T_{n}$ and the magnetoconductance $G_{MR}=G_{Q}%
\sum\nolimits_{n}(|t_{n,\uparrow}^{S}|^{2}-|t_{n,\downarrow}^{S}|^{2})$, with
$\uparrow(\downarrow)$ the majority(minority) spin direction in the $F$
electrode considered, $G_{Q}=e^{2}/h$, and $T_{n}=|t_{n,\uparrow}^{S}%
|^{2}+|t_{n,\downarrow}^{S}|^{2}$. In the case of a finite SDIPS, one must
also use the conductances $G_{\phi}^{F(S)}=2G_{Q}\sum\nolimits_{n}(\rho
_{n}^{F(S)}-4[\tau_{n}^{S(F)}/T_{n}])$, $G_{\xi}^{F(S)}=-G_{Q}\sum
\nolimits_{n}\tau_{n}^{S(F)}$ and $G_{\chi}^{F(S)}=G_{Q}\sum\nolimits_{n}%
T_{n}(\rho_{n}^{F(S)}+\tau_{n}^{S(F)})/4$, with $\rho_{n}^{m}%
=\operatorname{Im}[r_{n,\uparrow}^{m}r_{n,\downarrow}^{m~\ast}]$ and $\tau
_{n}^{m}=\operatorname{Im}[t_{n,\uparrow}^{m}t_{n,\downarrow}^{m~\ast}]$ for
$m\in\{S,F\}$. In the following, I will focus on the effects of $G_{\phi}^{F}$
and $G_{\phi}^{S}$, and I will assume $G_{MR}$, $G_{\xi}^{F(S)}$ and $G_{\chi
}^{F(S)}$ to be negligible, like found with a simple barrier model in the
limit $T_{n}\ll1$ and $E_{ex}\ll E_{F}$\cite{cottet05}. In this case, one
finds that the boundary conditions for the $S/F$ interface located at
$x=x_{j}=(-1)^{j}d_{s}/2$, with$\ j\in\{1,2\}$, write
\begin{equation}
\xi_{F}\left.  \frac{\partial\theta_{n,\sigma}^{F}}{\partial x}\right|
_{x_{j}}=(-1)^{j}\gamma_{T}\sin[\theta_{n,\sigma}^{F}-\theta_{n,\sigma}%
^{S}]+i\gamma_{\phi}^{F}\epsilon_{n,\sigma}^{\mathcal{C},j}\sin[\theta
_{n,\sigma}^{F}] \label{BCright}%
\end{equation}
and
\begin{equation}
\xi_{F}\left.  \frac{\partial\theta_{n,\sigma}^{F}}{\partial x}\right|
_{x_{j}}-\frac{\xi_{S}}{\gamma}\left.  \frac{\partial\theta_{n,\sigma}^{S}%
}{\partial x}\right|  _{x_{j}}=\sum\limits_{m\in\{F,S\}}i\gamma_{\phi}%
^{m}\epsilon_{n,\sigma}^{\mathcal{C},j}\sin[\theta_{n,\sigma}^{m}]
\label{BCleft}%
\end{equation}
where the indices $S$ and $F$ indicate whether $\theta_{n,\sigma}$ and its
derivative are taken at the $S$ or $F$ side of the interface. These equations
involve the reduced conductances $\gamma_{T}=G_{T}\xi_{F}/A\sigma_{F}$ and
$\gamma_{\phi}^{F(S)}=G_{\phi}^{F(S)}\xi_{F}/A\sigma_{F}$, the barrier
asymmetry coefficient $\gamma=\xi_{S}\sigma_{F}/\xi_{F}\sigma_{S}$, the
superconducting coherence lengthscale $\xi_{S}=(\hbar D_{S}/2\Delta
_{BCS})^{1/2}$, the magnetic coherence lengthscale $\xi_{F}=(\hbar
D_{F}/\left|  E_{ex}\right|  )^{1/2}$, the gap $\Delta_{BCS}$ for a bulk $S$,
the normal state conductivity $\sigma_{F(S)}$ of the $F(S)$ material, and the
junction area $A$. The coefficient $\epsilon_{n,j}^{\mathcal{C}}$ takes into
account the direction of the ferromagnetic polarization of electrode $F_{j}$
in configuration $\mathcal{C}\in\{P,AP\}$. One can use the convention
$\epsilon_{n,\sigma}^{P,j}=(-1)^{j}\sigma\mathrm{sgn}(\omega_{n})$ and
$\epsilon_{n,\sigma}^{AP,j}=\sigma\mathrm{sgn}(\omega_{n})$, in which the
factor $\mathrm{sgn}(\omega_{n})$ arising from the definition of
$\theta_{n,\sigma}$ and the terms $(-1)^{j}$ and $\sigma$ arising from the
boundary conditions have been included for compactness of the expressions.
Note that in the presence of a finite SDIPS i.e. $\gamma_{\phi}^{F(S)}\neq0$,
the right hand side of equation (\ref{BCleft}) is not zero contrarily to what
found in the spin-degenerate case \cite{Kuprianov}. In the general case,
$\gamma_{\phi}^{F}$ and $\gamma_{\phi}^{S}$ are different (see e.g. Appendix
A). This implies that, with the present approximations, each interface is
characterized by three parameters: $\gamma_{T}$, $\gamma_{\phi}^{F}$ and
$\gamma_{\phi}^{S}$. For the sake of simplicity, symmetric $F/S/F$ trilayers
are considered, so that $\gamma_{T}$, $\gamma_{\phi}^{F}$ and $\gamma_{\phi
}^{S}$ are the same for the two $S/F$ interfaces.

Before working out the above system of equations, it is interesting to note
that the angle $\theta_{n,\sigma}$ calculated in the parallel configuration
$\mathcal{C}=P$ for $x>0$ also corresponds to the angle $\theta_{n,\sigma}$
expected for a $S/F$ bilayer consisting of a superconductor $S$ for
$0<x<d_{S}/2$, and a ferromagnetic electrode $F$ for $x\in\{d_{S}%
/2,d_{S}/2+d_{F}\}$ (Figure 1.b). In practice, using a $F/S/F$ geometry can
allow one to obtain more information on spin effects, as shown below.\bigskip

\section{Case of a thin superconductor and a weak proximity effect in $F$}

\subsection{Spatial variations of the pairing angle}

I will assume that the amplitude of the superconducting correlations in
$F_{1(2)}$ is weak, i.e. $\left|  \theta_{n,\sigma}\right|  \ll1$ for
$x\in\{-d_{S}/2-d_{F},-d_{S}/2\}$ and $x\in\{d_{S}/2,d_{S}/2+d_{F}\}$
(\textit{hypothesis 1) }so that one can develop the Usadel equation
(\ref{UsadelF2}) at first order in $\theta_{n,\sigma}$. This leads to
\begin{equation}
\frac{\partial^{2}\theta_{n,\sigma}}{\partial x^{2}}-\left(  \frac
{k_{n,\sigma}^{\mathcal{C},j}}{\xi_{F}}\right)  ^{2}\theta_{n,\sigma}=0
\label{UsadelF}%
\end{equation}
in the ferromagnet $F_{j}$, with $j\in\{1,2\}$, $k_{n,\sigma}^{\mathcal{C}%
,j}=(2[i\eta_{n,\sigma}^{\mathcal{C},j}+\left|  \omega_{n}/E_{ex}\right|
)])^{1/2}$ and $\eta_{n,\sigma}^{\mathcal{C},j}=(-1)^{j}\epsilon_{n,\sigma
}^{\mathcal{C},j}$. Combining Eqs. (\ref{derZero}) and (\ref{UsadelF}), one
finds in $F_{j}$%

\begin{equation}
\theta_{n,\sigma}(x)=\theta_{n,\sigma}^{F}(x_{j})\frac{\cosh\left(  \left[
x-(-1)^{j}\left(  d_{F}+\frac{d_{S}}{2}\right)  \right]  \frac{k_{n,\sigma
}^{\mathcal{C},j}}{\xi_{F}}\right)  }{\cosh\left(  d_{F}\frac{k_{n,\sigma
}^{\mathcal{C},j}}{\xi_{F}}\right)  } \label{theta}%
\end{equation}
This result together with boundary condition (\ref{BCright}) leads to
\begin{equation}
\theta_{n,\sigma}^{F}(x_{j})=\frac{\gamma_{T}\sin(\theta_{n,\sigma}^{S}%
(x_{j}))}{\gamma_{T}\cos(\theta_{n,\sigma}^{S}(x_{j}))+i\gamma_{\phi}^{F}%
\eta_{n,\sigma}^{\mathcal{C},j}+B_{n,\sigma}^{\mathcal{C},j}} \label{thetaF}%
\end{equation}
with $B_{n,\sigma}^{\mathcal{C},j}=k_{n,\sigma}^{\mathcal{C},j}\tanh
[d_{F}k_{n,\sigma}^{\mathcal{C},j}/\xi_{F}]$. This allows to rewrite the
boundary condition (\ref{BCleft}) in closed form with respect to
$\theta_{n,\sigma}^{S}$, i.e.
\begin{equation}
\xi_{S}\left.  \frac{\partial\theta_{n,\sigma}^{S}}{\partial x}\right|
_{x_{j}}=(-1)^{j+1}\mathcal{L}_{j,n,\sigma}^{\mathcal{C}}\sin[\theta
_{n,\sigma}^{S}(x_{j})]
\end{equation}
for the $S/F$ interface located at $x=x_{j}$, with
\begin{equation}
\frac{\mathcal{L}_{j,n,\sigma}^{\mathcal{C}}}{\gamma}=\frac{\gamma
_{T}(B_{n,\sigma}^{\mathcal{C},j}+i\gamma_{\phi}^{F}\eta_{n,\sigma
}^{\mathcal{C},j})}{\gamma_{T}\cos(\theta_{n,\sigma}^{S})+B_{n,\sigma
}^{\mathcal{C},j}+i\gamma_{\phi}^{F}\eta_{n,\sigma}^{\mathcal{C},j}}%
+i\gamma_{\phi}^{S}\eta_{n,\sigma}^{\mathcal{C},j}%
\end{equation}
In the following, I will assume $\left|  E_{ex}\right|  \gg\Delta_{BCS}$ like
in most experiments, so that $k_{n,\sigma}^{\mathcal{C},j}=1+i\eta_{n,\sigma
}^{\mathcal{C},j}$. I will also assume $d_{S}/\xi_{S}\leq1$, so that one can
use, for the $\mathcal{C}$ configuration and $-d_{S}/2<x<d_{S}/2$,
\begin{equation}
\theta_{n,\sigma}(x)=\widetilde{\theta}_{n,\sigma}^{\mathcal{C}}%
-\alpha_{n,\sigma}^{\mathcal{C}}(x/\xi_{S})-\beta_{n,\sigma}^{\mathcal{C}%
}(x/\xi_{S})^{2} \label{thetaQuad}%
\end{equation}
with $\left|  \theta_{n,\sigma}(x)-\widetilde{\theta}_{n,\sigma}^{\mathcal{C}%
}\right|  \ll1$ (\textit{hypothesis 2}). Note that although experiments are
often performed in the limit of thick superconducting layers $d_{S}>\xi_{S}$,
assuming $d_{S}\leq\xi_{S}$ is not unrealistic since one can obtain diffusive
superconducting layers with a thickness $d_{S}\sim\xi_{S}$ (see e.g.
Ref.~\onlinecite{Fominov}). Furthermore, using relatively low values of
$d_{S}$ is more favorable for obtaining efficient superconducting
spin-switches\cite{Baladie}. Hypothesis 2 allows one to develop $\sin
(\theta_{n,\sigma})$ and $\cos(\theta_{n,\sigma})$\ at first order with
respect to $\theta_{n,\sigma}-\widetilde{\theta}_{n,\sigma}^{\mathcal{C}}$ in
$S$. Accordingly, I will neglect the space-dependence of $\Delta(x)$ and
assign to it the value $\Delta^{\mathcal{C}}$ in configuration $\mathcal{C}$.
Note that I do not make any assumption on the value of the angle
$\widetilde{\theta}_{n,\sigma}^{\mathcal{C}}$, which is not necessarily close
to the bulk BCS value.\ The coefficient $\mathcal{L}_{j,n,\sigma}%
^{\mathcal{C}}$ is transformed into its conjugate when the magnetic
polarization of electrode $F_{j}$ is reversed. Therefore, I will note
$\mathcal{L}_{1,n,\sigma}^{P}=\mathcal{L}_{2,n,\sigma}^{P(AP)}=\mathcal{L}%
_{n}^{\sigma}$ and $\mathcal{L}_{1,n,\sigma}^{AP}=(\mathcal{L}_{n}^{\sigma
})^{\ast}$. The above assumptions lead to $\alpha_{n,\sigma}^{P}=0$,
\begin{equation}
\beta_{n,\sigma}^{P}=\frac{4\mathcal{L}_{n}^{\sigma}\sin(\widetilde{\theta
}_{n,\sigma}^{\mathcal{C}})}{4\delta_{S}+\cos(\widetilde{\theta}_{n,\sigma
}^{\mathcal{C}})\delta_{S}^{2}\mathcal{L}_{n}^{\sigma}} \label{bP}%
\end{equation}%
\begin{equation}
\alpha_{n,\sigma}^{AP}=i\operatorname{Im}[\mathcal{L}_{n}^{\sigma}]\frac
{4\sin(\widetilde{\theta}_{n,\sigma}^{\mathcal{C}})-\cos(\widetilde{\theta
}_{n,\sigma}^{\mathcal{C}})\delta_{S}^{2}\beta_{AP}}{4+2\operatorname{Re}%
[\mathcal{L}_{n}^{\sigma}]\cos(\widetilde{\theta}_{n,\sigma}^{\mathcal{C}%
})\delta_{S}}%
\end{equation}
and
\begin{equation}
\beta_{n,\sigma}^{AP}=\frac{\left(  8\operatorname{Re}[\mathcal{L}_{n}%
^{\sigma}]+4\left|  \mathcal{L}_{n}^{\sigma}\right|  ^{2}\cos(\widetilde
{\theta}_{n,\sigma}^{\mathcal{C}})\delta_{S}\right)  \sin(\widetilde{\theta
}_{n,\sigma}^{\mathcal{C}})}{8\delta_{S}+6\operatorname{Re}[\mathcal{L}%
_{n}^{\sigma}]\cos(\widetilde{\theta}_{n,\sigma}^{\mathcal{C}})\delta_{S}%
^{2}+\left|  \mathcal{L}_{n}^{\sigma}\right|  ^{2}\cos^{2}(\widetilde{\theta
}_{n,\sigma}^{\mathcal{C}})\delta_{S}^{3}} \label{bAP}%
\end{equation}
with $\delta_{S}=d_{S}/\xi_{S}$. On the other hand, from (\ref{UsadelS}), one
finds
\begin{equation}
\beta_{n,\sigma}^{\mathcal{C}}=\frac{\Delta^{\mathcal{C}}\cos(\widetilde
{\theta}_{n,\sigma}^{\mathcal{C}})-\left|  \omega_{n}\right|  \sin
(\widetilde{\theta}_{n,\sigma}^{\mathcal{C}})}{2\Delta_{BCS}} \label{bc}%
\end{equation}
The comparison between equations (\ref{bP}), (\ref{bAP}) and (\ref{bc}) allows
one to find $\widetilde{\theta}_{n,\sigma}^{\mathcal{C}}$ as a function of
$\Delta^{\mathcal{C}}$. Then, one has to calculate $\Delta^{\mathcal{C}}$ by
using the self-consistency relation (\ref{Delta}). I will study below the DOS
and the critical temperature following from these Eqs., in a limit which leads
to simple analytical expressions.

\subsection{Low-temperature density of states of $S/F$ heterostructures}

The DOS of the ferromagnets $F_{1}$ and $F_{2}$ of Figure 1.a can be probed at
$x=\pm(d_{F}+d_{S}/2)$ by performing tunnelling spectroscopy through an
insulating layer. So far, this quantity has been less
measured\cite{TakisN,Kontos04,Courtois,Reymond} than critical temperatures or
supercurrents. However, this way of probing the superconducting proximity
effect is very interesting because it allows one to obtain spectroscopic
information. It has been shown that the zero-energy DOS of a $F$ layer in
contact with a superconductor oscillates with the thickness of $F$. For
certain thicknesses, this zero-energy DOS can even become higher than its
normal state value $N_{0}$ \cite{theoDOS,Zareyan,Bergeret,yokohama}, as shown
experimentally in Ref.~\onlinecite{TakisN}. Remarkably, the SDIPS can shift
these oscillations\cite{cottet05}. Although the energy dependence of the DOS
of diffusive $S/F$ structures has raised some theoretical and experimental
interest, the effect of the SDIPS on this energy dependence has not been
investigated so far.

\begin{figure}[ptb]
\includegraphics[width=0.8\linewidth]{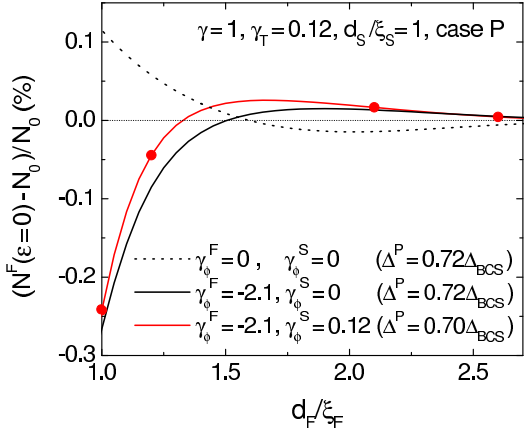}\caption{Zero energy density of
states $N^{F}(\varepsilon=0)$ at $x=d_{F}+d_{S}/2$ as a function of $d_{F}$,
for the $F/S/F$ structure of Figure 1.a in the $P$ configuration, with
$\gamma=1$, $\gamma_{T}=0.12$, and $d_{S}/\xi_{S}=1$. The different curves
correspond to different SDIPS parameters:$\ \gamma_{\phi}^{F}=\gamma_{\phi
}^{S}=0$ (black dotted curve), $\{\gamma_{\phi}^{F}=-2.1$, $\gamma_{\phi}%
^{S}=0\}$ (black full curve), and $\{\gamma_{\phi}^{F}=-2.1$, $\gamma_{\phi
}^{S}=0.12\}$ (red full curve). The low-temperature self-consistent gap
$\Delta^{P}$\ found for these three different cases is indicated at the bottom
right of the Figure. Using a strong value for $\gamma_{\phi}^{F}$ allows one
to change significantly the phase of the oscillations of $N_{\sigma}^{F}(0)$
with $d_{F}$ (effect similar to Ref. \onlinecite
{cottet05}). One can check that the hypotheses 1 to 4 are valid for the
parameters used in this Figure. Note that for the different cases considered
here, the critical temperature $T_{c}^{P}$ of the structure (see Sec. III.F)
is such that $0.615T_{c}^{BCS}<T_{c}^{P}<0.635T_{c}^{BCS}$. The four red
points correspond to the red full curves shown in Fig. \ref{DOS2}.}%
\label{DOS1}%
\end{figure}

For calculating analytically the low-temperature DOS of the structure of Fig.
1.a., one can assume $\gamma_{T}^{2}\cos(\theta_{n,\sigma}^{S}(x_{j}))/\left|
(\right.  \gamma_{T}\cos(\theta_{n,\sigma}^{S}(x_{j}))+i\gamma_{\phi}^{F}%
\eta_{n,\sigma}^{P,2}+B_{n,\sigma}^{P,2})(\gamma_{T}+i\gamma_{\phi}^{S}%
\eta_{n,\sigma}^{P,2}\left.  )\right|  \ll1$ (\textit{hypothesis 3}), which
leads to $\mathcal{L}_{n}^{\sigma}=\gamma\left(  \gamma_{T}+i\gamma_{\phi}%
^{S}\sigma\mathrm{sgn}(\omega_{n})\right)  $. This hypothesis is e.g. valid
for $d_{F}\geq\xi_{F}$ and any value of $\gamma_{\phi}^{S(F)}$ if $\gamma_{T}$
is relatively small (see e.g. Fig. \ref{DOS1}). I will also assume that the
lowest order terms in $\delta_{S}$ prevail in the numerators and denominators
of expressions (\ref{bP}) and (\ref{bAP}), i.e. $\beta_{n,\sigma}^{P}%
\sim\mathcal{L}_{n}^{\sigma}\sin(\widetilde{\theta}_{n,\sigma}^{\mathcal{C}%
})\delta_{S}^{-1}$ and $\beta_{n,\sigma}^{AP}\sim\operatorname{Re}%
[\mathcal{L}_{n}^{\sigma}]\sin(\widetilde{\theta}_{n,\sigma}^{\mathcal{C}%
})\delta_{S}^{-1}$ (\textit{hypothesis 4}). Taking into account hypothesis 3
and $\gamma\sim1$, hypothesis 4 is valid provided $\gamma_{T}$ and $\left|
\gamma_{\phi}^{S}\right|  $ are relatively small compared to $1$. Importantly,
hypotheses 3 and 4 are less restrictive regarding the value of $\gamma_{\phi
}^{F}$. Accordingly, I will often use $\gamma_{T},\left|  \gamma_{\phi}%
^{S}\right|  \ll\left|  \gamma_{\phi}^{F}\right|  $ in the Figs. of this
paper. Hypotheses 3 and 4 lead to
\begin{equation}
\widetilde{\theta}_{n,\sigma}^{\mathcal{C}}=\mathrm{\arctan}\left(
\frac{\Delta^{\mathcal{C}}}{\left|  \omega_{n}\right|  +\Omega_{n,\sigma
}^{\mathcal{C}}}\right)  \label{thetaS}%
\end{equation}
with
\begin{equation}
\Omega_{n,\sigma}^{P}=2\Delta_{BCS}\gamma(\gamma_{T}+i\gamma_{\phi}^{S}%
\sigma\mathrm{sgn}(\omega_{n}))\delta_{S}^{-1} \label{omegaa}%
\end{equation}
and
\begin{equation}
\Omega_{n,\sigma}^{AP}=2\Delta_{BCS}\gamma\gamma_{T}\delta_{S}^{-1}
\label{Om2}%
\end{equation}
Equations (\ref{thetaS}) and (\ref{omegaa}) show that an effective
magnetic field $H_{eff}$ appears in the $S$ layer in the $P$ configuration,
due to $\gamma_{\phi}^{S}\neq0$. From Eqs. (\ref{thetaS}) and (\ref{omegaa}),
$H_{eff}$ can be expressed as
\begin{equation}
g\mu_{B}H_{eff}=\frac{\hbar v_{F}^{S}}{d_{s}}\frac{2G_{\phi}^{S}\ell_{e}^{S}%
}{3\sigma_{S}A}=2E_{TH}^{S}\frac{G_{\phi}^{S}}{G_{S}} \label{heff}%
\end{equation}
Here, $v_{F}^{S}$ and $\ell_{e}^{S}$ denote the Fermi velocity and mean free
path in $S$, and $G_{S}=\sigma_{S}A/d_{S}$ and $E_{TH}^{S}=\hbar D_{S}%
/d_{S}^{2}$ denote the normal state conductance and the Thouless energy of the
$S$ layer of Fig. 1.a. From Equations (\ref{thetaS}) and (\ref{Om2}), the
effective field effect disappears in the $AP$ configuration because the two
contacts are assumed to be symmetric, and therefore, their contributions to
$H_{eff}$ compensate each other in the $AP$ case. Note that, in principle, the
$\gamma_{\phi}^{F}$ term can induce an effective field analogue to $H_{eff}$
in the $F$ layer, but this effect is not relevant in the regime studied in
this paper (see Appendix B). The effects of $H_{eff}$ on the DOS of the
structure will be investigated in next paragraphs. In order to calculate
$\Delta^{\mathcal{C}}$, one has to combine the self-consistency relation
(\ref{Delta}) with Eq. (\ref{thetaS}), which gives, at low temperatures,
\begin{equation}
\operatorname{Re}[\log(\frac{\Omega_{n,\sigma}^{\mathcal{C}}+\sqrt{\left(
\Omega_{n,\sigma}^{\mathcal{C}}\right)  ^{2}+\left(  \Delta^{\mathcal{C}%
}\right)  ^{2}}}{\Delta_{BCS}})]=0 \label{DeltaEq}%
\end{equation}
This equation can be solved numerically. The resulting $\Delta^{\mathcal{C}}$
is independent from the values of $n$ and $\sigma$ used in Eq. (\ref{DeltaEq}%
). Then, the value of the pairing angle $\theta_{n,\sigma}$ in the
ferromagnets can be found by using Eqs. (\ref{theta}), (\ref{thetaF}),
(\ref{thetaQuad}) and (\ref{thetaS}). Note that for $\gamma\ll1$, the above
Eqs. are in agreement with formula (5) of Ref.~\onlinecite{cottet05}, obtained
with rigid boundary conditions, i.e. $\theta_{n,\sigma}$ equal to its bulk
value at the $S$ side. The energy dependence of $\theta_{n,\sigma}$ can be
found by performing the analytic continuation $\omega_{n}=-i\varepsilon
+\Gamma$ and $\mathrm{sgn}(\omega_{n})=1$ in the above equations. The rate
$\Gamma=0.05$ is used to account for inelastic processes \cite{theseW}. At
last, the density of states $N_{\sigma}(x,\varepsilon)$ at position $x$ for
the spin direction $\sigma\in\{\uparrow,\downarrow\}$ can be calculated by
using $N_{\sigma}(x,\varepsilon)=\left(  N_{0}/2\right)  \operatorname{Re}%
[\cos[\theta_{n,\sigma}(x)]]$, where $N_{0}/2$ is the normal density of states
per spin direction.

\begin{figure}[ptb]
\includegraphics[width=0.85\linewidth]{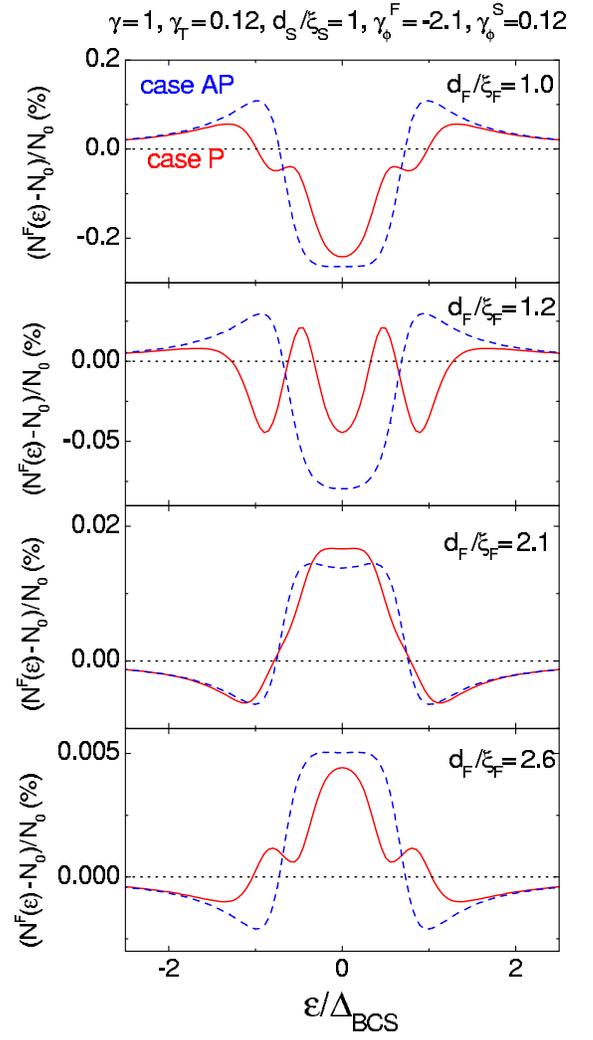}\caption{Density of states
$N^{F}(\varepsilon)$ at $x=d_{F}+d_{S}/2$ as a function of the energy
$\varepsilon$, for the $F/S/F$ structure of Figure 1.a with $\gamma=1$,
$\gamma_{T}=0.12$, $d_{S}/\xi_{S}=1$, $\gamma_{\phi}^{F}=-2.1$ and
$\gamma_{\phi}^{S}=0.12$. The different panels correspond to different values
of $d_{F}/\xi_{F}$. The red full curves correspond to the $P$ configuration
and the blue dashed curves to the $AP$ configuration. For finite values of
$\gamma_{\phi}^{S}$, some characteristic ''double-gap''\ structures appear in
$N_{\sigma}^{F}(\varepsilon)$ in the $P$ case, due to a SDIPS-induced
effective field appearing in $S$. The visibility of this effect strongly
depends on the thickness $d_{F}$ of the $F$ layers. For the different cases
considered here, one can check that hypotheses 1 to 4 are valid and that the
critical temperature $T_{c}^{\mathcal{C}}$ of the structure (see Sec. III.F)
is such that $0.62T_{c}^{BCS}<T_{c}^{\mathcal{C}}<0.67T_{c}^{BCS}$. Note that
for the above parameters and $\gamma_{\phi}^{S}=0$, the curves obtained in the
$P$ and $AP$ case would be identical and very close to the $AP$ curves shown
here. }%
\label{DOS2}%
\end{figure}

In the following, I will mainly focus on $N^{F}(\varepsilon)=\sum
\nolimits_{\sigma\in\{\uparrow,\downarrow\}}N_{\sigma}(x=d_{F},\varepsilon)$.
Figure \ref{DOS1} shows the variations of the zero energy density of states
$N^{F}(\varepsilon=0)$ as a function of $d_{F}$, for interface parameters
$\gamma_{T}=\gamma_{\phi}^{S}=0.12$ and $\gamma_{\phi}^{F}=-2.1$. Importantly,
the value $\gamma_{T}=0.12$ seems realistic, at least for the weakly polarized
\textrm{Nb/Pd}$_{0.9}$\textrm{Ni}$_{0.1}$ bilayers used in Ref. \onlinecite
{TakisN}, for which one finds $\gamma_{T}\sim0.15$ (see Ref. \onlinecite
{expl}). In addition, a simple barrier model suggests that the situation
$\left|  G_{\phi}^{F}\right|  \gg G_{T}$, with $G_{\phi}^{F}<0$ can happen
(see Appendix A). The value $\gamma_{\phi}^{F}=-2.1$ used in Fig. \ref{DOS1}
thus seems possible\textbf{. }One can see that $\gamma_{\phi}^{F}$ can change
significantly the phase of the oscillations of $N^{F}(\varepsilon=0)$ with
$d_{F}$ (this effect has already been studied in Ref.~\onlinecite{cottet05} in
the case of rigid boundary conditions but I recall it here for the sake of
completeness). In Fig. \ref{DOS1}, using for $\gamma_{\phi}^{F}$ a strong
negative value allows to get $N^{F}(0)<N_{0}$ for the lowest values of $d_{F}%
$, like often found in experiments. Note that from Fig. \ref{DOS1},
$\gamma_{\phi}^{S}$ can also shift the oscillations of $N^{F}(\varepsilon=0)$
with $d_{F}$. Here, this effect is much weaker than that of $\gamma_{\phi}%
^{F}$, but one has to keep in mind that the limit $\gamma_{\phi}^{S}\ll
\gamma_{\phi}^{F}$ is considered.

Equation (\ref{heff}) shows that a measurement of $H_{eff}$ should allow to
determine the conductance $G_{\phi}^{S}$ of a diffusive $S/F$ interface. In
this context, studying the energy dependence of $N^{F}(\varepsilon)$ is very
interesting, because it can allow to see clear signatures of $H_{eff}$, as
shown below. Figure \ref{DOS2} shows the energy dependence of $N^{F}%
(\varepsilon)$ in the $P$ and $AP$ configurations, for a finite value of
$\gamma_{\phi}^{S}$ and different values of $d_{F}$. For $\mathcal{C}=P$,
$N^{F}(\varepsilon)$ shows some ''double-gap''\ structures which disappear if
the device is switched to the $AP$ configuration. These double structures are
an indirect manifestation of the effective magnetic field $H_{eff}$ occurring
in $S$ in the $P$ configuration, due to $\gamma_{\phi}^{S}\neq0$. Although
$H_{eff}$ is localized in the $S$ layer, the double-gap structure that this
field produces in the DOS of $S$ is transmitted to the DOS of $F$ due to
Andreev reflections occurring at the $S/F$ interfaces, as shown by Eq.
(\ref{thetaF}). Interestingly, Rowell and McMillan have already observed that
an internal property of a $S$ layer can be seen through the superconducting
proximity effect occurring in a nearby normal layer. More precisely, these
authors have found that the DOS\ of an \textrm{Ag} layer can reveal the phonon
spectrum of an adjacent superconducting \textrm{Pb} layer\cite{Rowell}.
Remarkably, the visibility of $H_{eff}$ in $N^{F}(\varepsilon)$ is modulated
by quantum interferences occurring in $F$. Indeed, $H_{eff}$ is more visible
for certain values of $d_{F}$ (e.g. $d_{F}/\xi_{F}=1.0$ or $1.2$ in Fig.
\ref{DOS2}) than others (e.g. $d_{F}/\xi_{F}=2.1$ in Fig. \ref{DOS2}), due to
the $d_{F}$-dependence of Eq.(\ref{thetaF}).

\begin{figure}[ptb]
\includegraphics[width=1\linewidth]{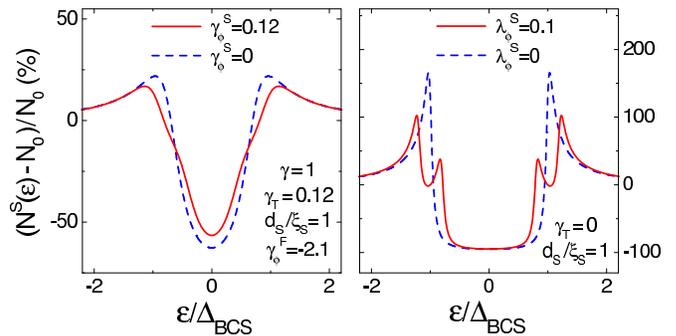}\caption{Energy dependent density
of states $N^{S}(\varepsilon)$ measurable at the S side ($x=0$) of the $S/F$
structure of Figure 1.b. The left panel corresponds to the case of a metallic
$F$ contact with parameters corresponding to that of Figs. \ref{DOS1} and
\ref{DOS2}. In this case, the DOS measured at the $S$ side does not allow to
resolve the SDIPS-induced effective field $H_{eff}$ better than a DOS
measurement at the $F$ side of the structure, as can be seen from a comparison
with Fig. \ref{DOS2}. For certain values of $d_{F}$, $H_{eff}$ is even more
visible in $N^{F}(\varepsilon)$ than $N^{S}(\varepsilon)$. The right panel
corresponds to the case in which $F$ is not a metal but an insulating
ferromagnet, i.e. $\gamma_{T}=0$ . In this case, one can use the reduced SDIPS
parameter $\lambda_{\phi}^{S}=G_{\phi}^{S}\xi_{S}/A\sigma_{S}$. For
$\lambda_{\phi}^{S}\neq0$, $N^{S}(\varepsilon)$ shows strong signatures of the
effective field $H_{eff}$ $\ $induced by the $FI$ layer in $S$. One can check
that the hypotheses 1 to 4 are valid for the parameters used in this Figure.}%
\label{DOS3}%
\end{figure}

It is useful to note that the SDIPS-induced effective field $H_{eff}$ should
also occur in the $S/F$ bilayer of Figure 1.b. In this case, the Thouless
energy and normal state conductance of the $S$ layer correspond to
$\widetilde{E}_{TH}^{S}=4E_{TH}^{S}$ and $\widetilde{G}_{S}=2G_{S}$
respectively, so that one finds $g\mu_{B}H_{eff}=\widetilde{E}_{TH}^{S}%
G_{\phi}^{S}/\widetilde{G}_{S}$. Double gap structures strikingly similar to
those shown in Figure \ref{DOS2} were indeed measured very recently by P.
SanGiorgio et al., at the ferromagnetic side of diffusive \textrm{Nb/Ni}
bilayers, in the absence of any external magnetic field\cite{SanGiorgio}.
Remarkably, the visibility of the observed double structures varies with
$d_{F}$, as predicted above. Note that in $S/F$ bilayers, the field $H_{eff}$
should also be observable directly at the $S$ side by measuring $N^{S}%
(\varepsilon)=\sum\nolimits_{\sigma\in\{\uparrow,\downarrow\}}N_{\sigma
}(x=0,\varepsilon)$. However, for parameters comparable to those of Figure
\ref{DOS2}, this should not enhance the resolution on $H_{eff}$ (see figure
\ref{DOS3}, left). For certain values of $d_{F}$, $H_{eff}$ is even more
visible in $N^{F}(\varepsilon)$ than in $N^{S}(\varepsilon)$(see Figure
\ref{DOS2}).

Before concluding this section, I would like to emphasize that from Eqs.
(\ref{thetaS}) and (\ref{Om2}), in the $AP$ configuration, the SDIPS-induced
effective field $H_{eff}$ disappears for the $F/S/F$ structure considered in
this paper because the two contacts are assumed to be symmetric and have thus
opposite contributions to $H_{eff}$ in the $AP$ case. In the case of a
dissymmetric structure, this should not be true anymore, but the SDIPS-induced
effective field should nevertheless vary from the $P$ to the $AP$ case. This
is one practical advantage of working with $F/S/F$ trilayers instead of $S/F$ bilayers.

\subsection{Low-temperature density of states of $S/FI$ bilayers}

Twenty years ago, internal Zeeman fields were observed in superconducting Al
layers contacted to different types of ferromagnetic insulators ($FI$) (see
Refs.~\onlinecite{Tedrow,Meservey,MooderaTheoDiff,Hao}). Using a
\textit{ballistic} $S/FI$ bilayer model, Ref.~\onlinecite
{Tokuyasu} suggested that the observed internal fields could be induced by the
SDIPS\cite{Tokuyasu}. However, the inadequacy of this ballistic approach for
modeling the actual experiments was pointed out in Ref.~\onlinecite
{Hao}. Most of the experiments on Al$/FI$ interfaces were interpreted by their
authors in terms of a diffusive approach with no SDIPS, and an internal Zeeman
field added arbitrarily in the Al layer (see Refs. \onlinecite
{MooderaTheoDiff,Hao,Alexander}). The calculations of Section III.B. provide a
microscopic justification for the use of such an internal field in the
diffusive model. Indeed, using $\gamma_{T}=0$ in the above calculations allows
one to address the case of diffusive $S/FI$ bilayers. One finds that the
SDIPS-induced effective field $H_{eff}$ of Eq. (\ref{heff}) can occur in a
thin diffusive $S$ layer contacted to a $FI$ layer. This effective field
effect can be seen e.g. in the density of states $N^{S}(\varepsilon)$ of the
$S$ layer at $x=0$ (see Figure \ref{DOS3}, right). Remarkably, it was found
experimentally\cite{Hao} that $H_{eff}$ scales with $d_{s}^{-1}$, in agreement
with Eq. (\ref{heff})\cite{NoteDG}.

\subsection{SDIPS-induced effective fields in other types of system}

Interestingly, the SDIPS can induce effective field effects in other types of
systems. First, the case of $S/N/FI$ trilayers with a thickness $d_{N}$ of
normal metal $N$ has been studied theoretically\cite{demoBC,applications}. In
this case, a conductance $G_{\phi}^{N}$ similar to $G_{\phi}^{S}$ can be
introduced to take into account the SDIPS for electrons reflected by the $FI$
layer. The $N$ layer is subject to an effective field $g\mu_{B}H_{eff}%
^{\prime}=E_{TH}G_{\phi}^{N}/G_{N}$ with $G_{N}$ the conductance of $N$. The
expression of $H_{eff}^{\prime}$ is analogue to that of $H_{eff}$ (see Eq.
\ref{heff}), up to a factor 2\ which accounts for the symmetry of the $F/S/F$
structure with respect to $x=0$ in the $P$ configuration. Secondly, an
effective field $H_{eff}^{^{\prime\prime}}$ defined by $g\mu_{B}%
H_{eff}^{^{\prime\prime}}=(\hbar v_{F}^{W}/2L)(\varphi_{L}^{\uparrow}%
+\varphi_{R}^{\uparrow}-\varphi_{L}^{\downarrow}-\varphi_{R}^{\downarrow})$
has been predicted for a resonant single-channel ballistic wire with length
$L$ placed between two ferromagnetic contacts\cite{CottetEurophys}, with
$\varphi_{L(R)}^{\sigma}$ the reflection phase of electrons with spin $\sigma$
incident from the wire onto the left(right) contact, and $v_{F}^{W}$ is the
Fermi velocity in the wire. The expression of $H_{eff}^{^{\prime\prime}}$ also
shows strong similarities with that of $H_{eff}$ (see Eq. \ref{heff}, middle
term). In practice, signatures of the field $H_{eff}^{^{\prime\prime}}$ could
be identified in a carbon nanotube contacted with two ferromagnetic contacts
\cite{Cottet06,Sahoo}. The fields $H_{eff}$, $H_{eff}^{^{\prime}}$ and
$H_{eff}^{^{\prime\prime}}$\ have the same physical origin: the energies of
the states localized in the central conductor of the structure depend on spin
due to the spin-dependent phase shifts acquired by electrons at the boundaries
of this conductor. In all cases, the DOS of the central conductor reveals the
existence of the SDIPS-induced effective field only if it already presents a
strong energy dependence near the Fermi energy in the absence of a SDIPS. In
the $F/S/F$ case, this energy dependence is provided by the existence of the
superconducting gap in $S$\cite{cond}. In the $S/N/FI $ case, it is provided
by the existence of a superconducting minigap in $N$. At last, in the case of
the ballistic wire, it is provided by the existence of resonant states in the wire.

\subsection{Comparison between the present work and other models for data
interpretation in $S/F$ heterostructures}

For characterizing the properties of $S/F$ interfaces, one has to interpret
the experimental data showing the oscillations of the density of states
$N^{F}(\varepsilon)$ in $F$ with the thickness $d_{F}$ of $F$ (or the
oscillations of the critical current $I_{0}$ of a $S/F/S$ Josephson with
$d_{F}$). However, if one uses a simple description with spin-degenerate
boundary conditions, the amplitude and the phase of these signals are not
independent, which makes the agreement with experimental results impossible in
most cases. The SDIPS concept can solve this problem since it produces a shift
of the signals oscillations with respect to the $G_{\phi}^{S(F)}=0$ case.
However, in many cases, the observed shifts were attributed to the existence
of a magnetically dead layer (MDL) at the $F$ side of the $S/F$ interface (see
e.g. Refs. ~\onlinecite
{TakisN,Oboznov,Weides}). In other cases, the discrepancy between the theory
and the data was solved by taking into account spin-scattering processes in
the $F$ layer (see e.g. Ref. \onlinecite
{SellierPRB}). In order to have a better insight into superconducting
proximity effect experiments, one must stress the importance of estimating
experimentally the MDL thickness and the spin-scattering rate. In principle,
spin-scattering rates can be estimated experimentally, as was done for
instance for the \textrm{CuNi} alloy\cite{Pratt} which is frequently used in
proximity effect measurements (see e.g. \onlinecite
{SellierPRB,Oboznov}). An experimental determination of the MDL thickness has
also been performed in a few structures used to measure $T_{c}$ or $I_{0}%
$\cite{Muhge,Aarts,Piano2,Piano,Bell}, but, so far, this parameter has not
been used for a real quantitative analysis of the data. In some situations, a
model combining the SDIPS with spin-scattering and/or a MDL\ may be necessary.
In any case, it is important to point out that descriptions based on
spin-degenerate boundary conditions are, in principle, incomplete since they
do not account for the effective field effect described in section III.B.

Before concluding this section, it is interesting to note that the effective
field effect produced by $G_{\phi}^{S}$ in $S$ or the phase shift of the
spatial oscillations of the DOS provided by $G_{\phi}^{F}$ in $F$ will remain
qualitatively similar when the signs of $G_{\phi}^{S}$ and $G_{\phi}^{F}$ are
changed (not shown). The sign of $G_{\phi}^{F}$ in the experiments of
Refs.~\onlinecite
{TakisN,TakisI} could be determined from a quantitative comparison between the
theory and the data\cite{cottet05}. Below, we present a study of the critical
temperature of $S/F$ structures which can give information on the signs of
$G_{\phi}^{S}$ and $G_{\phi}^{F}$ through qualitative signatures.

\subsection{Critical temperature of $S/F$ heterostructures}

The critical temperature of $S/F$ hybrid structures has already been the topic
of many theoretical (see e.g. Refs.~\onlinecite
{DeGennes,TcFSFth,Baladie,TcSFth,You}) and experimental (see e.g.
Refs.~\onlinecite{TcSF,SSspinswitch}) studies, but the effects of the SDIPS on
this quantity have raised little attention so far. I show below that the SDIPS
can significantly modify the critical temperatures of $S/F$ diffusive
structures. Calculating the critical temperature $T_{c}^{\mathcal{C}}$ of the
structure of Figure 1.a in configuration $\mathcal{C}$ requires to consider
the limit in which superconducting correlations are weak in $S$ as well as in
$F$ (hypotheses 1 and 2 are then automatically satisfied). Equations
(\ref{bP}), (\ref{bAP}) and (\ref{bc}), then lead to

\begin{figure}[ptb]
\includegraphics[width=1\linewidth]{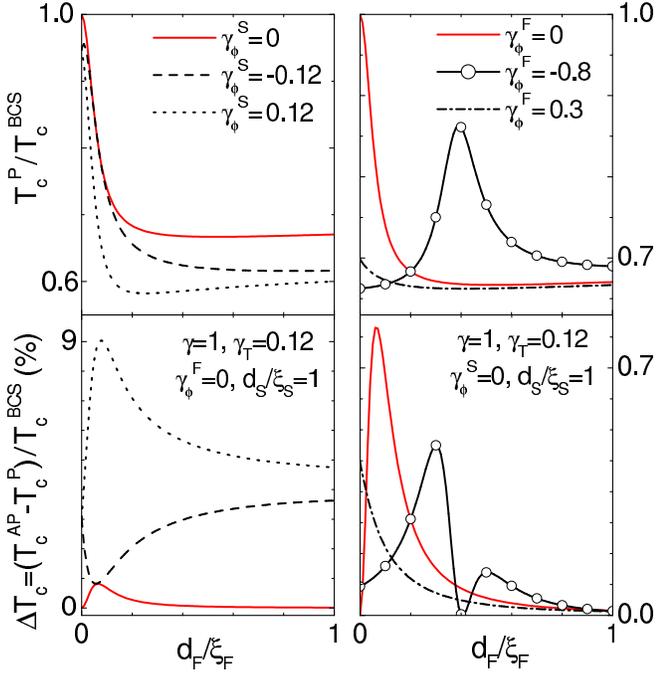}\caption{Critical temperature
$T_{c}^{P}$ for the $F/S/F$ structure of Figure 1.a in the $P$ configuration
(top panels) and difference $\Delta T_{c}=T_{c}^{AP}-T_{c}^{P}$ between the
critical temperatures in the $P$ and $AP$ configurations (bottom panels), as a
function of the thickness $d_{F}$ of the $F$ layers. The left panels show the
effect of a finite $\gamma_{\phi}^{S}$ and the right panels the effect of a
finite $\gamma_{\phi}^{F}$. The four panels show the case $\gamma_{\phi}%
^{S}=\gamma_{\phi}^{F}=0$ with full red lines, for comparison. The other
parameters used in the Figure are $\gamma=1$, $\gamma_{T}=0.12$ and $d_{S}%
/\xi_{S}=1$. The SDIPS modifies $T_{c}^{\mathcal{C}}$ and $\Delta T_{c}$ in a
quantitative or qualitative way, depending on the case considered. The type of
effects produced by the SDIPS on $T_{c}^{\mathcal{C}}$ and $\Delta T_{c}$
strongly depend on the signs of $\gamma_{\phi}^{F}$ and $\gamma_{\phi}^{S}$.}%
\label{FSFTc}%
\end{figure}%
\begin{equation}
\widetilde{\theta}_{n,\sigma}^{\mathcal{C}}=\frac{\Delta^{\mathcal{C}}%
}{\left|  \omega_{n}\right|  +2\Delta_{BCS}b_{n,\sigma}^{\mathcal{C}}%
\delta_{S}^{-1}}%
\end{equation}
with
\begin{equation}
b_{n,\sigma}^{P}=\frac{4\mathcal{L}_{n,\sigma}^{0}}{4+\delta_{S}%
\mathcal{L}_{n,\sigma}^{0}} \label{bp}%
\end{equation}%
\begin{equation}
b_{n,\sigma}^{AP}=\frac{8\operatorname{Re}[\mathcal{L}_{n,\sigma}%
^{0}]+4\left|  \mathcal{L}_{n,\sigma}^{0}\right|  ^{2}\delta_{S}%
}{8+6\operatorname{Re}[\mathcal{L}_{n,\sigma}^{0}]\delta_{S}+\left|
\mathcal{L}_{n,\sigma}^{0}\right|  ^{2}\delta_{S}^{2}} \label{bap}%
\end{equation}
and
\begin{equation}
\frac{\mathcal{L}_{n,\sigma}^{0}}{\gamma}=\frac{\gamma_{T}(B_{n,\sigma}%
^{P,2}+i\gamma_{\phi}^{F}\sigma\mathrm{sgn}(\omega_{n}))}{\gamma
_{T}+B_{n,\sigma}^{P,2}+i\gamma_{\phi}^{F}\sigma\mathrm{sgn}(\omega_{n}%
)}+i\gamma_{\phi}^{S}\sigma\mathrm{sgn}(\omega_{n})
\end{equation}
These Eqs. together with (\ref{Delta}) lead to
\begin{equation}
\log\left(  \frac{T_{c}^{BCS}}{T_{c}^{\mathcal{C}}}\right)  =\operatorname{Re}%
\left[  \Psi\left(  \frac{1}{2}+\frac{b_{n,\sigma}^{\mathcal{C}}}{\delta
_{S}\exp(\Gamma)}\frac{T_{c}^{BCS}}{T_{c}^{\mathcal{C}}}\right)  \right]
-\Psi\left(  \frac{1}{2}\right)  \label{Tc}%
\end{equation}
where $\Gamma$ denotes Euler's constant. The resulting $T_{c}^{\mathcal{C}}$
is independent from the values of $n$ and $\sigma$ used in Eq. (\ref{Tc}).
Note that in the case $\gamma_{\phi}^{S}=\gamma_{\phi}^{F}=0$ and $\delta
_{S}\rightarrow0$, this equation is in agreement with Eqs. (18) and (19) of
Ref.~\onlinecite{Baladie}.

Performing a numerical resolution of Eq. (\ref{Tc}) together with (\ref{bp})
and (\ref{bap}), one obtains the results of Fig. \ref{FSFTc}, which shows the
critical temperature $T_{c}^{P}$ of the structure in the $P$ configuration
(top panels) and the difference $\Delta T_{c}=T_{c}^{AP}-T_{c}^{P}$ (bottom
panels) as a function of $d_{F}$, for different interface
parameters\cite{approx}. Here, for simplicity, I consider cases where
$T_{c}^{\mathcal{C}}$ does not show a strongly reentrant behavior, i.e. a
cancellation of $T_{c}^{\mathcal{C}}$ in a certain interval of $d_{F}$. Such a
behavior can happen for instance for larger values of $\gamma_{T}$ (see e.g.
Ref.~\onlinecite{Baladie}) or $\gamma_{\phi}^{(S)F}$, but this corresponds to
a minority\cite{Tccoupe} of the experimental observations made so far. The
four panels of Fig. \ref{FSFTc} show the case $\gamma_{\phi}^{S}=\gamma_{\phi
}^{F}=0$ with full red lines, for comparison with the other cases, where the
SDIPS is finite. I first comment the results obtained for the $T_{c}^{P}%
(d_{F})$ curves (top panels of Fig. \ref{FSFTc}). In some cases, the
SDIPS\ can modify quantitatively the $T_{c}^{P}(d_{F})$ curves, for instance
by amplifying the dip expected in $T_{c}^{P}(d_{F})$ in the absence of a SDIPS
(see e.g. dotted curve in the upper left panel, corresponding to $\gamma
_{\phi}^{S}>0$). In some other cases the SDIPS can modify qualitatively the
$T_{c}^{P}(d_{F})$ curves, for instance by transforming the minimum expected
in $T_{c}^{P}(d_{F})$ into a maximum\ (see e.g. curve with circles in the
upper right panel, corresponding to $\gamma_{\phi}^{F}<0$). I now comment the
results obtained for the $\Delta T_{c}(d_{F})$ curves (bottom panels of Fig.
\ref{FSFTc}). In the range of parameters studied in this work, one always
finds $\Delta T_{c}>0$ because the effects of the two $F$ layers on $S$ are
partially compensated in the $AP$ case. In the absence of a SDIPS and for a
low $\gamma_{T}$, the $\Delta T_{c}(d_{F})$ curve presents a maximum at a
finite value of $d_{F}$ (see red full lines in bottom panels). In some cases,
the SDIPS\ can modify quantitatively the $\Delta T_{c}(d_{F})$ curves, for
instance by increasing the value of this maximum (see dotted curve in the
bottom left panel, corresponding to $\gamma_{\phi}^{S}>0$). In other cases,
the SDIPS can modify qualitatively the $\Delta T_{c}(d_{F})$ curves, for
instance by turning the maximum expected in $\Delta T_{c}(d_{F})$ into a
minimum (see dashed curve in the bottom left panel, corresponding to
$\gamma_{\phi}^{S}<0$), or by increasing the complexity of the variations of
$\Delta T_{c}$ with $d_{F}$ (see curve with circles in the bottom right panel,
corresponding to $\gamma_{\phi}^{F}<0$), or by transforming $\Delta
T_{c}(d_{F})$ into a monotonically decreasing curve with a reduced amplitude
(see dot-dashed curve in bottom right panel, corresponding to $\gamma_{\phi
}^{F}>0$). Remarkably, the type of effects produced by the SDIPS on the
$T_{c}^{\mathcal{C}}(d_{F})$ and $\Delta T_{c}(d_{F})$ curves depend on the
sign of $\gamma_{\phi}^{F}$ and $\gamma_{\phi}^{S}$. Critical temperature
measurements can thus be an interesting way to determine the values of
$\gamma_{\phi}^{F}$ and $\gamma_{\phi}^{S}$. For simplicity, I have shown here
results for $\gamma_{\phi}^{F}\neq0$ and $\gamma_{\phi}^{S}\neq0$ separately.
Nevertheless, the behavior predicted for $\gamma_{\phi}^{F}\neq0$ and
$\gamma_{\phi}^{S}\neq0$ simultaneously remain highly informative on the
values of $\gamma_{\phi}^{F}$ and $\gamma_{\phi}^{S}$. Note that if
$\gamma_{\phi}^{F}$ and $\gamma_{\phi}^{S}$ are increased compared to the
values used in Fig. \ref{FSFTc}, the $T_{c}^{\mathcal{C}}(d_{F})$ curve gets
some cancellation points like in Ref.~\onlinecite{Baladie}, but the behaviors
of $T_{c}^{\mathcal{C}}$ and $\Delta T_{c}$ remain, in many cases,
qualitatively dependent on the signs of $\gamma_{\phi}^{F}$ and $\gamma_{\phi
}^{S}$ (not shown).

Before concluding, I note that with the approximations used in the previous
section (III.B), the electronic correlations inside $S$ were affected by the
presence of the $F$ electrodes through the parameter $\gamma_{\phi}^{S}$ only.
Consequently, the self-consistent gap $\Delta^{\mathcal{C}}$ of the $S$ layer
was independent from $d_{F}$, and it was furthermore identical for the $P$ and
$AP$ configurations for $\gamma_{\phi}^{S}=0$. In the present section, I have
not neglected the dependence of $T_{c}^{P}$ and $T_{c}^{AP}$ on $d_{F}$ and
$\gamma_{\phi}^{F}$ because I have considered parameters for which hypothesis
3 is not acceptable anymore. In particular, I have used lower values for
$d_{F}$\cite{regimes}.

\section{Conclusion}

This article shows that the Spin-Dependence of Interfacial Phase Shifts
(SDIPS) can have a large variety of signatures in diffusive
superconducting/ferromagnetic ($S/F$) heterostructures. Ref. \onlinecite
{cottet05} had already predicted that the SDIPS produces a phase shifting of
the oscillations of the superconducting correlations with the thickness of $F
$ layers. This article shows that this is not the only consequence of the
SDIPS in $S/F$ circuits. In particular, the SDIPS can produce an effective
magnetic field in a diffusive $S$ layer contacted to a diffusive $F$ layer.
This effective field can be seen e.g. through the DOS of the diffusive $F$
layer, with a visibility which oscillates with the thickness of $F$. The SDIPS
can also modify significantly the variations of the critical temperature of a
$S/F$ bilayer or a $F/S/F$ trilayer with the thickness of $F $, either in a
quantitative or in a qualitative way, depending on the regime of parameters
considered and in particular the sign of the conductances $G_{\phi}^{S}$ and
$G_{\phi}^{F}$ used to account for the SDIPS of the $S/F$ interfaces. In the
case of a $F/S/F$ spin valve, this last result also holds for the
thickness-dependence of the difference between the critical temperatures in
the parallel and antiparallel lead configurations. These effects should help
to determine the parameters $G_{\phi}^{S}$ and $G_{\phi}^{F}$ of diffusive
$S/F$ interfaces. The calculations shown in this paper are also appropriate to
the case of thin diffusive $S$ layers contacted to ferromagnetic insulators.

\textit{I thank P. SanGiorgio for showing me his experimental data prior to
publication, which stimulated part III.B of this work. I acknowledge
discussions with T. Kontos and W. Belzig. This work was supported by grants
from the Swiss National Science Foundation and R\'{e}gion Ile-de-France.}

\section{Appendix A: Parameters of a $S/F$ interface from a Dirac barrier model}

\begin{figure}[h]
\includegraphics[width=0.85\linewidth]{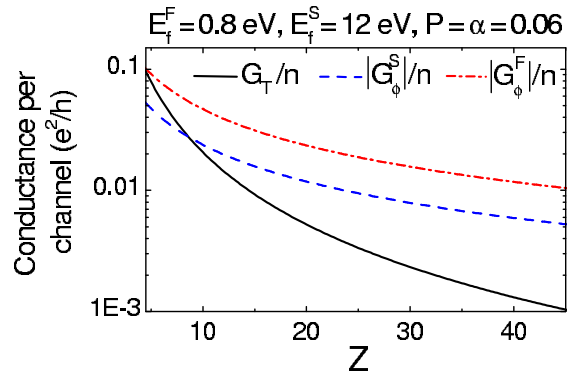}\caption{Conductances $G_{T}$
(full line), $\left|  G_{\phi}^{S}\right|  $ (dashed lines), and $\left|
G_{\phi}^{F}\right|  $ (dash-dotted lines) reduced by the number of channels
$n$, as a function of the spin-averaged barrier strength $Z$ of a $S/F$
interface modeled with a Dirac barrier (see text). The curves were obtained
with typical Fermi energies $E_{f}^{F}=0.8~\mathrm{eV}$ and $E_{f}%
^{S}=12~\mathrm{eV}$ in $F$ and $S$ respectively, a spin-polarization $P=0.06$
of the density of states in $F$, and a spin asymmetry $\alpha=0.06$ for the
barrier. Note that for the set of parameters used this Figure, one finds
$G_{\phi}^{F}<0$ and $G_{\phi}^{S}<0$. However, for other parameters (e.g. by
using $\alpha<0$), one can reverse the signs of $G_{\phi}^{F}$ and $G_{\phi
}^{S}$, or obtain opposite signs for $G_{\phi}^{F}$ and $G_{\phi}^{S}$ (not
shown).}%
\label{Dirac}%
\end{figure}

The exact values of the conductances $G_{T}$, $G_{\phi}^{F}$ and $G_{\phi}%
^{S}$ of a $S/F$ interface depend on the details of this interface and on the
microscopic structure of the contacted materials. Nevertheless, it is already
interesting to study a simplified Dirac barrier model which shows that the
parameters regime assumed in this paper is, in principle, possible. Neglecting
the transverse part of the electrons motion, one finds $r_{n,\sigma}%
^{S(F)}=(k_{S(F)}^{\sigma}-k_{F(S)}^{\sigma}-iZ^{\sigma})/D^{\sigma}$ and
$t_{n,\sigma}^{S(F)}=2(k_{S}^{\sigma}k_{F}^{\sigma})^{1/2}/D^{\sigma}$ with
$D^{\sigma}=k_{S}^{\sigma}+k_{F}^{\sigma}+iZ^{\sigma}$. Here, $k_{S(F)}%
^{\sigma}$ is the Fermi electronic wavevector in $S(F)$ and $Z^{\sigma}$ is
the strength of the Dirac barrier for electrons with spin $\sigma$. I use
$\hbar k_{S}^{\sigma}=(2m_{e}E_{f}^{S})^{1/2}$, with $m_{e}$ the free electron
mass and $E_{f}^{S}$ the Fermi level in $S$. For $F$, I use an s-band Stoner
model, in which $\hbar k_{F}^{\sigma}=(2m_{e}E_{f}^{F}[1\pm2P/(1+P^{2}%
)])^{1/2}$, with $P$ the spin-polarization of the density of states in $F$ and
$E_{f}^{F}$ the Fermi level in $F$. I assume that the barrier can have a spin
dependence $\alpha=(Z^{\downarrow}-Z^{\uparrow})/(Z^{\uparrow}+Z^{\downarrow
})$ due to the ferromagnetic contact material used to form the interface. The
results given by this approach are shown in Fig. \ref{Dirac}. The conductances
$G_{T}$, $\left|  G_{\phi}^{F}\right|  $ and $\left|  G_{\phi}^{S}\right|  $,
reduced by the number of channels $n$, are shown as a function of the
spin-averaged barrier strength $Z=(Z^{\uparrow}+Z^{\downarrow})/2$. The three
types of conductances go to zero when $Z$ goes to infinity with $\alpha$
constant\cite{lim}. One can see that in the limit $T_{n}\ll1$ (i.e. here
$G_{T}h/ne^{2}\ll1$) and $P\ll1$ in which the boundary conditions
(\ref{BCright},\ref{BCleft}) have been derived, $\left|  G_{\phi}^{F}\right|
$ can be significantly stronger than $G_{T}$, as assumed in Figs. \ref{DOS1}
and \ref{DOS2}. Note that for the set of parameters used in Fig. \ref{Dirac},
one finds $G_{\phi}^{F}<0$ and $G_{\phi}^{S}<0$. However, for other parameters
(e.g. by using $\alpha<0$), one can reverse the signs of $G_{\phi}^{F}$ and
$G_{\phi}^{S}$, or obtain opposite signs for $G_{\phi}^{F}$ and $G_{\phi}^{S}$
(not shown). This suggests that there is no fundamental constraint on the
signs of $G_{\phi}^{F}$ and $G_{\phi}^{S}$ in the general case.

\section{Appendix B: Effective field effect in a $F$ layer}

This appendix reconsiders the case of the $S/F$ bilayer of Fig.1.b, in the
limit $d_{F}\ll\xi_{F}$ where $\theta_{n,\sigma}(x)$ can be approximated with
a quadratic form in the $F$ layer. From Eqs. (\ref{UsadelF2}) and
(\ref{BCright}), one finds, for $x\in\lbrack d_{S}/2,d_{S}/2+d_{F}]$,%

\begin{equation}
\theta_{n,\sigma}(x)=\widetilde{\theta}_{n,\sigma}^{F}+b_{n,\sigma}^{F}\left[
\frac{d_{S}-2x}{d_{F}}+\left(  \frac{d_{S}-2x}{2d_{F}}\right)  ^{2}\right]
\end{equation}
with%

\begin{equation}
\widetilde{\theta}_{n,\sigma}^{F}=\arctan\left(  \frac{\gamma_{T}%
\sin(\widetilde{\theta}_{n,\sigma}^{S})}{\gamma_{T}\cos(\widetilde{\theta
}_{n,\sigma}^{S})+\frac{2\xi_{F}}{d_{F}}\frac{\Omega_{n}}{E_{TH}^{F}}%
+i\gamma_{\phi}^{F}\eta_{n,\sigma}^{P,2}}\right)  \label{ThetaF}%
\end{equation}%
\begin{equation}
b_{n,\sigma}^{F}=\frac{\Omega_{n}\sin(\widetilde{\theta}_{n,\sigma}^{F}%
)}{E_{TH}^{F}}%
\end{equation}
$\Omega_{n}=\left|  \omega_{n}\right|  +iE_{ex}\eta_{n,\sigma}^{P,2}$ and
$\widetilde{\theta}_{n,\sigma}^{S}=\theta_{n,\sigma}^{S}(x=d_{S}/2)$. Here,
$G_{F}=\sigma_{F}A/d_{F}$ and $E_{TH}^{F}=\hbar D_{F}/d_{F}^{2}$ denote the
conductance and Thouless energy of the $F$ layer (note that $d_{F}\ll\xi
_{F}\Longleftrightarrow E_{ex}\ll E_{TH}^{F}$). From completeness, I also give
the analogous equations for $x\in\lbrack0,d_{S}/2]$. Assuming that
$\theta_{n,\sigma}(x)$ can be approximated with a quadratic form in the $S$
layer, one finds, from Eqs. (\ref{UsadelS}) and (\ref{BCleft}),
\begin{equation}
\theta_{n,\sigma}(x)=\widetilde{\theta}_{n,\sigma}^{S}+b_{n,\sigma}^{S}\left[
\frac{2(2x-d_{S})}{d_{S}}+\left(  \frac{2x-d_{S}}{d_{S}}\right)  ^{2}\right]
\end{equation}
with
\begin{equation}
\widetilde{\theta}_{n,\sigma}^{S}=\mathrm{\arctan}\left(  \frac{\gamma
\gamma_{T}\sin(\widetilde{\theta}_{n,\sigma}^{F})+\frac{4\xi_{S}}{d_{S}}%
\frac{\Delta}{\widetilde{E}_{TH}^{S}}}{\gamma\gamma_{T}\cos(\widetilde{\theta
}_{n,\sigma}^{F})+\frac{4\xi_{S}}{d_{S}}\frac{\left|  \omega_{n}\right|
}{\widetilde{E}_{TH}^{S}}+i\gamma\gamma_{\phi}^{S}\eta_{n,\sigma}^{P,2}%
}\right)  \label{ThetaS}%
\end{equation}
and
\begin{equation}
b_{n,\sigma}^{S}=\frac{\left|  \omega_{n}\right|  \sin(\widetilde{\theta
}_{n,\sigma}^{S})-\Delta\cos(\widetilde{\theta}_{n,\sigma}^{S})}{\widetilde
{E}_{TH}^{S}}%
\end{equation}
The notations used in the above equations are the same as in Section III. In
particular, the Thouless energy and the normal state conductance of the $S$
layer with thickness $d_{S}/2$ are denoted $\widetilde{G}_{S}=2\sigma
_{S}A/d_{S}$ and $\widetilde{E}_{TH}^{S}=4\hbar D_{S}/d_{S}^{2}$, and one has
$\eta_{n,\sigma}^{P,2}=\sigma\mathrm{sgn}(\omega_{n})$. In the limit
$\widetilde{\theta}_{n,\sigma}^{F}\ll1$, Eq. (\ref{ThetaS}) is in agreement
with Eqs. (\ref{thetaS}) and (\ref{omegaa}). The analytic continuation of Eq.
(\ref{ThetaS}) shows that the $S$ layer is subject to the effective field
$H_{eff}=\widetilde{E}_{TH}^{S}G_{\phi}^{S}/\widetilde{G}_{S}$, in agreement
with Eq. (\ref{heff}). The analytic continuation of Eq. (\ref{ThetaF}) shows
that the $F$ layer is subject to an analogue effective field $H_{eff}^{F}$,
defined by
\begin{equation}
g\mu_{B}H_{eff}^{F}=E_{TH}^{F}\frac{G_{\phi}^{F}}{G_{F}}%
\end{equation}
It is interesting to compare Eqs. (\ref{ThetaF}) and (\ref{thetaF}). First,
note that in the regime $\left|  \widetilde{\theta}_{n,\sigma}^{F}\right|
\ll1$, Eq. (\ref{ThetaF}) can be recovered from the low $d_{F}$-limit of Eq.
(\ref{thetaF}). The interest of Eq. (\ref{ThetaF}) is that it is valid beyond
the regime of a weak superconducting proximity effect studied in part III. In
particular, Eq. (\ref{ThetaF}) indicates that, in principle, the field
$E_{TH}^{F}$ can occur in a thin $F$ layer for any value of the tunneling
conductance $G_{T}$ of the $S/F$ interface. Interestingly, in the case of a
thick $F$ layer $d_{F}\geq\xi_{F}/2$, from Eq. (\ref{thetaF}), the
contribution of $G_{\phi}^{F}$ to the pairing angle $\theta_{n,\sigma}^{F}(x)$
cannot be put anymore under the form of an effective exchange field, due to
the non-linearity of the $B_{n,\sigma}^{\mathcal{C},j}$ term. Thus, strictly
speaking, the concept of a SDIPS-induced effective field is valid only in the
limit of thin metallic layers. However, it might be possible, in principle, to
observe reminiscences of the double-gap structure appearing in $N^{F}%
(\varepsilon)$ in the regime of intermediate layer thicknesses $d_{F}\sim
\xi_{F}$ at least, due to the continuity of the equations. Then, one can
wonder why this effect does not appear in section III. This is due to the
regime of parameters chosen: section III assumes $E_{ex}\gg\Delta_{BCS}$ like
in most experiments, and it furthermore focuses on the typical energy range
probed in superconducting proximity effect measurements, i.e. $\left|
\varepsilon\right|  \lesssim2\Delta_{BCS}$. In such a regime, one can neglect
the term $\left|  \omega_{n}\right|  $ compared to $iE_{ex}\sigma
\mathrm{sgn}(\omega_{n})$ in Eqs. (\ref{thetaF}) and (\ref{ThetaF}), so that
$H_{eff}^{F}$ does not emerge in the model. Accordingly, signatures of the
SDIPS-induced effective field $H_{eff}^{F}$ are not likely to be observed in practice.


\begin{thebibliography}{99}
\bibitem{NoteDG}The fact that $H_{eff}$ scales with $d_{s}^{-1}$ is also in
agreement with Eq. (1) of Ref.~\onlinecite
{DeGennes}, proposed by De Gennes from Cooper's argument\cite{Cooper}.
\bibitem{cond}Without superconductivity, $H_{eff}$ would not be not resolvable
anymore in the $S/F$ structures considered in this paper, since it was assumed
implicitly that the normal-state DOS\ of these structures is constant near the
Fermi level.
\bibitem{approx}With the parameters of Fig. \ref{FSFTc}, the approximations
$b_{n,\sigma}^{P}=\mathcal{L}_{n,\sigma}^{0}$ and $b_{n,\sigma}^{AP}%
=\operatorname{Re}[\mathcal{L}_{n,\sigma}^{0}]$ used in Ref.~\onlinecite
{Baladie} would be sufficient for calculating the $T_{c}^{P}$ and $T_{c}^{AP}$
curves, but not the difference $\Delta T_{c}$. Indeed, for the different cases
studied in this Figure, the relative difference between the results obtained
with these approximations and those given by Eqs. (\ref{bp}) and (\ref{bap})
is smaller than $\sim0.8\%$ for $T_{c}^{P}$ and $T_{c}^{AP}$, but it can reach
$\sim18\%$ for $\Delta T_{c}$. This is why I have used the general equations
(\ref{bp}) and (\ref{bap}) for plotting Figure \ref{FSFTc}.
\bibitem{regimes}Note that both regimes $d_{F}\leq\xi_{F}$ (see e.g. Ref
\onlinecite{Courtois}) and $d_{F}\geq\xi_{F}$ (see e.g. Ref.~\onlinecite{You})
seem to be accessible in practice.
\end{thebibliography}
\end{document}